\documentstyle[12pt]{article}
\def \LSP{\widetilde{\chi_1}^0}
\def \N2{\widetilde{\chi_2}^0}
\def \CH{\widetilde{\chi}^{\pm}}
\def \W1{\widetilde{\chi}_1^{\pm}}   
\def \WC2{\widetilde{\chi}_2^{\pm}}
\def \SNU{\tilde{\nu}}
\def \BARSNU{\tilde{\bar{\nu}}}

\def \CH{\widetilde{\chi}^{\pm}}
\def \W1{\widetilde{\chi}_1^{\pm}}
\def \WC2{\widetilde{\chi}_2^{\pm}}
\def \SNU{\tilde{\nu}}
\def \BARSNU{\tilde{\bar{\nu}}}

\def \MLSP{m_{\widetilde{\chi_1}^0}}
\def \MN2{m_{\widetilde{\chi_2}^0}}
\def \MCH{m_{\widetilde{\chi}^{\pm}}}
\def \MSNU{m_{\tilde{\nu}}}

\begin{document}
\setcounter{page}{0}
\thispagestyle{empty}
\begin{center}
{\large\bf  ON THE OBSERVABILITY OF "INVISIBLE" / "NEARLY INVISIBLE" CHARGINOS}\\
\bigskip
{\normalsize Amitava Datta{\footnote {adatta@juphys.ernet.in}}
 and Shyamapada Maity{\footnote {shyama@juphys.ernet.in}}}\\
{\footnotesize 
Department of Physics, Jadavpur University, Calcutta 700 032,
India.}\\
\end{center}
\vskip 5pt

\begin{center}
{\bf ABSTRACT }
\end{center}
{\footnotesize
It is shown that if the charginos decay into very soft leptons or hadrons +
$\not\!\!{E}$ due to  degeneracy/ near- degeneracy with the LSP or
 the sneutrino, the observability of the recently proposed signal 
via the single photon (+ soft particles) + $\not\!\!{E}$ 
channel crucially depends on the magnitude of the $\SNU$
mass due to  destructive interferences in the matrix element squared. If the
$\SNU$'s and, consequently, left-sleptons are relatively light, the size
of the signal, previously computed in the limit $\MSNU \rightarrow \infty$
only, is drastically reduced. We present the formula for the signal cross 
section in a model independent way and discuss the observability of the
signal at LEP 192 and NLC energies.} 

\vskip 10pt
\newpage
\section
{Introduction}
The search for Supersymmetry (SUSY) \cite{1}, the most attractive candidate 
for physics beyond the Standard Model (SM), is a programme with top
priority in present day high energy experiments. From the non-observation 
of the sparticles at various colliders only  lower limit on their masses have
been obtained so far.

These limits are, however, derived with certain caveats. For example
it is now well known that the most stringent lower limit on the
lighter chargino ($\CH$) mass \cite{2,3} 
 is obtained from  LEP under the assumption that
they are significantly heavier than the lightest supersymmetric
particle (LSP) ($\LSP$). Typically the strongest limit on the
lighter chargino mass $(\MCH)$ is obtained for ${\Delta m}\geq$ 10
GeV, where ${{\Delta m} \equiv \MCH - \MLSP}$. If ${\Delta m}$ is
smaller, the particles arising from the decay $ \CH \rightarrow \LSP +
X$, where $ X $ is any hadronic or leptonic state, become rather
soft. Triggering on such particles could be extremely difficult if
$\Delta m \ll$ 10 GeV. Even if the problem of triggering could be
overcome, the signal is likely to be swamped by the large  SM
background from two photon processes, which also involve soft final
state particles. It is therefore not impossible that charginos are
within the current striking range of  LEP but their presence is
not revealed due to unexpected degeneracies (or near degeneracies) in
the sparticle spectrum. Looking for such charginos is indeed of
crucial importance since they may happen to be the only visible
sparticles within the kinematical reach of the machine concerned.

Although such a specific mass pattern occurs in a relatively tiny
region of the large parameter space of the Minimal Supersymmetric
Standared Model (MSSM), search strategies for these apparently
less probable scenarios should nevertheless be chalked out. This is so
in view of the tremendous importance of discovering  SUSY or
ruling it out without any shreds of doubt. Moreover, the so called
improbable scenarios quite often look sensible once various
uncertainties due to GUT scale or Planck scale physics are properly
taken into account, as will be briefly discussed in the next section.

In some recent works \cite{4,5} the search strategies for detceting
invisible / nearly invisible charginos due to extreme/approximate
degeneracy between $\CH$ and $\LSP$ were elaborated (to be reviewed in
some detail in the next section). The essential point is to observe
the reaction
\begin{equation}
e^+e^- \rightarrow \tilde{\chi}^+ \tilde{\chi}^- \gamma  
\end{equation}
where the event can be triggered by the hard photon. If the charginos
are practically invisible such events may show up as an excess of
single photon + $\not\!\!{E} $(missing energy) events over the SM
background at  LEP 192  or  NLC, provided the chargino mass
happens to be in a limited range which depends on the c.m. energy. On
the other hand if the decay products are soft but visible and/or a
short track corresponding to the charginos can be observed, the signal
becomes virtually background free. In this case much larger charginos
masses (almost up to the kinematic limit) can be probed \cite{4,5}.

The cross section for process (1) was calculated in \cite{4,5}. This
was done under the assumption that the sneutrino ${\tilde{(\nu)}}$
exchange diagrams in the t-channel (Fig. 1) are negligible (i.e.
${m_{\tilde{\nu}} \rightarrow \infty}$). While this assumption may be
valid in the context of specfic models (see the next section for the
details) \cite{6} discussed by these authors, it is not true in
general. On the contrary one can easily construct models with small
${\Delta m}$ but arbitrary sneutrino masses (to be elaborated in the
next section). We have, therefore, computed the full cross-section
taking into account all the diagrams of Fig. 1 and their
interferences. Our model independent calculation reveals that the
assumption $\MSNU\rightarrow\infty$ is not conservative. The
cross-section in question is in fact significantly reduced for smaller
$\MSNU$ due to destructive interferences between s and t channel
diagrams. This is reminiscent of a similar cancellation in the case of
chargino pair production without an accompanying photon.

It may be argued that if the sneutrinos and, consequently, the
left-sleptons are indeed light then SUSY is likely to be discovered
through the slepton channel. Thus the chargino signals may not be of
crucial importance in this scenario.  Our model independent
calculations reveal that even if the sneutrinos are assumed to be of
finite mass but significantly heavier than the charginos (in fact
$\MSNU$ may be outside the kinematic reach of the
collider), the cross section may still be affected appreciably.
Whether the right-sleptons are within the kinematic reach in this case
is of course a model dependent issue which will be briefly taken up in
the next section.

We also take this opportunity to point out that the charginos may very
well be invisible or nearly invisible due to their degeneracy with the
sneutrinos. This happens if ${m_{\tilde\nu} < m_{\tilde\chi^\pm}}$,
with ${\Delta m^{\prime} \equiv m_{\tilde\chi^\pm} -m_{\tilde\nu}}$ rather
small. Under this circumstances the left - sleptons are necessarily
heavier than the chargino and right - slepton masses are anyway of
little consequence for chargino decays.

In this case the charginos dominantly decay into ${\tilde\chi^\pm
\rightarrow l\tilde\nu_l}$ and the sneutrinos decay into the invisible
channel ${\tilde\nu \rightarrow \nu\tilde\chi_1^0}$. Here the leptons
in the final state may be soft due to small ${\Delta
m^{\prime}}$. That the detection of the charginos may indeed be
problematic due to this degeneracy has also been noted by the 
 LEP collaborations \cite{2,3}.  We emphasize that unlike the small
${\Delta m}$ scenario, a small ${\Delta m^{\prime}}$ can be easily
accommodated in the popular  SUSYGUTS with gaugino mass
unification at the  GUT scale.

It is obvious that process (1) may turn out to be useful in this case
as well. However, as we will see the cross-section depends very
sensetively on ${m_{\tilde\nu}}$ and a full calculation as is done in
this paper is required.

The plan of the paper is as follows. In section {\bf 2} we shall
review the scenarios with small ${\Delta m}$ and point out that some
of them may naturally have small ${m_{\tilde\nu}}$. We shall then
briefly review the present experimental limits on
${m_{\tilde\chi^\pm}}$ and their sensitivity to ${\Delta m}$. A
similar review of the theoretical and experimental situations for
small ${\Delta m^{\prime}}$ will also be presented.

In section {\bf 3} the results for the cross-section of process (1)
(the relevant formulae are given in the appendix) will be presented as
a function ${m_{\tilde\nu}}$ and the agreement of our results with
those of \cite{4,5} in the limit ${m_{\tilde\nu}\rightarrow\infty
}$ will be demonstrated. The feasibility of discovering
invisible/nearly invisible charginos at  LEP 192  and  NLC will
then be commented upon. The summary of our results and conclusions
will be presented in section {\bf 4}.

\section{\bf The scenarios with small ${\Delta m /\Delta m^{\prime}}$ and  SUSY
search} 

In the most popular versions of the  MSSM an universal gaugino mass 
at the  GUT scale ${M_G}$ is assumed.  This implies 
\begin{equation}
 M_1 = M_2 = M_3 = M_{1/2}
\end{equation} 
at ${M_G}$ where ${M_1,M_2,M_3}$ are the masses of the  U(1),
 SU(2), and  SU(3) gauginos respectively. In obtaining the
${M_i}$'s at the energy scale of experimental interest the standard
renormalisation group (RG) equations \cite{7} are employed. This
yields reasonably large ${\Delta m}$ as long as $\mu$ is appreciably
large compared to ${M_1}$ and ${M_2}$. However if $|\mu| \ll M_1 ,M_2$
then $\CH,\LSP $ and the second lightest neutralino ($\N2$) turn out
to be higgsino like and approximately degenerate with masses $\sim
\mu$. Moreover, since $\CH $ is higgsino like the diagram with
sneutrino exchange in the t-channel  Fig.1 contributes
negligibly. Hence the cross-section presented in \cite{4,5} with
$\MSNU \rightarrow\infty$ are valid. In our subsequent discussions we
shall not consider this case in detail.  Howerever, as already
remarked in \cite{4,5} this scenario is not very natural if one is
interested in practically invisible charginos corresponding to
extremely small $\Delta m$. In order to give some rough estimates,
 we note that if $M_2 \geq
5$ TeV, $M_1$ determined by the condition (2) and 50 GeV $\leq
|\mu|\leq$ 100 GeV, then $\Delta m \leq 1$ GeV. For charginos nearly
degenerate with the  LSP, the choice of $\Delta m$ need not be
necessarily unnatural. For example, $\Delta m \approx $ 5 GeV can be
accommodated with $M_2\approx$1 TeV, 50 GeV $\leq|\mu|\leq $ 100 GeV.
As noted in \cite{5a}, loop corrections to this mass splitting may turn
out to be very important in this case and should be taken into account
for more refined results.

For larger $\mu$, howerever, one must allow for significant departures
from Eq. (2) in order to get a small $\Delta m$. For example, it was
argued in \cite{4,5} that if $M_2 < M_1$ at the weak scale and
$|\mu|\gg M_1 , M_2$, then $\CH$ and $\LSP$ are both wino like and
closely degenerate with $\MLSP \sim \MCH \sim M_2 $. The possiblity of
non-universal gaugino masses and their phenomenological implications
have been considered in the recent literature \cite{8}. In the most
general case, however, the scalar masses, though dependent on the
gaugino masses through the RG Eq.s, receive additional contributions
from a free parameter $m_0$, the common scalar mass at the  GUT
scale. The sneutrino mass, therefore, can be very large compared to
the electroweak gaugino masses or comparable to them depending on the
choice of $m_0$. The approximation $\MSNU \rightarrow \infty$ used in
\cite{4,5} is therefore model dependent. Of course in specific models
small $\Delta m$ and large $\MSNU$ can be accommodated
simultaneously. This, for example, is the case in the O-II model
\cite{6} in which both slepton and squark masses turn out to be much
larger than the gaugino masses.

There are, however, models which can accommodate relatively light
sneutrinos and small $\Delta m$. In certain models gaugino masses are
generated by a chiral superfield that appears linearly in the gauge
kinetic function and its auxilary F component acquires a vacuum
expectation value (VEV). In such a scenario the gaugino masses are
given by \cite{8}.
\begin{equation} 
{\cal{L}}_m \sim {\langle F \rangle_{ab}\over{M_{Planck}}} 
{\lambda}^{a}{\lambda}^{b}
\end{equation}
where ${\lambda}^{a,b}$ are the gaugino fields. If the theory above
the  GUT scale is described by an  SU(5) model, then F can
belong to any irreducible represention which appears in the symmetric
product of two adjoints:
\begin{equation}
(24
\otimes 24) = 1 \oplus 24\oplus 75\oplus 200
\end{equation}
where only 1 leads to universal masses at $M_G$. Requiring a
nonvanishing VEV only for that component of F which is neutral with
respect to the  SM gauge group, one obtains $\langle F
\rangle_{ab} = C_{a}{\delta}_{ab}$ where the calculable coefficient
$C_a$ determine the relative magnitudes of gauginos masses at
$M_G$. As noted in \cite{8}, for the choice 200 in Eq. 4 a small
$\Delta m$ results at the weak scale with both $\CH$ and $\LSP$ being
gaugino like. In this case $\MSNU$ is not necessarily large (its
magnitude depends on $m_0$). Thus the impact of the sneutrino exchange
diagram in Fig. 1 is worth studying (see the next section for
numerical results).

In several recent works the interesting phenomenological consequences of 
scenarios with $\MSNU<\MCH$ have been discussed  \cite{9,10,11}.

In such a scenario the sneutrinos decay via the invisible mode
$\SNU\rightarrow\nu\LSP$. The charginos on the other hand decay via
$\CH\rightarrow\ l\SNU$ with 100\% branching ratio. Relatively light
sneutrinos can be accommodated in conventional N=1 SUGRA type models
with common scalar and gaugino masses at the GUT scale
\cite{10,11} albeit in a narrow region of the parameter space. 
They however, become much more probable if the
possiblity of non-universal scalar masses in various SUGRA based
models are taken into account \cite{11}.

If in addition $\Delta m^{\prime} = \MCH - \MSNU$ happens to be small,
this scenario will lead to invisible/nearly invisible charginos as
discussed in the introduction. Of course such degeneracy or near
degeneracy appeares to be somewhat accidental. Yet in our opinion this
case is worth studying since it  leaves open the posiblity of unconventional
chargino signals even with the popular boundary condition of
universal gaugino masses (see Eq. 2). It is obvious that in this case
a full calculation of the cross-section taking all diagrams of Fig. 1
into account, as is done in the next section, is imperative. If the
charginos are invisible then the nummber of $\gamma + \not\!\!{E}$ events which
are already large in this scenario due to invisible sneutrinos
\cite{10} will be further enhanced. If on the other hand the charginos
are nearly invisible background free events as discussed in
 \cite{4,5} will be obtained from process (1).

The experimental reports on charginos searches at  LEP also
corroborates the difficulties in observing the chargino signal for
small $\Delta m$ or $\Delta m^{\prime}$ \cite{2,3}. The  OPAL
collaboration has recently reported results of chargino neutralino
searches \cite{3} at $\sqrt{s}$ = 170 and 172 GeV at  LEP. From
\cite{3}, it is seen that (see Fig 15 of \cite{3}) roughly
speaking a given chargino mass can not be excluded if $\Delta m
\leq$ 5 GeV. This point is also clear from Table-VIII of
\cite{3} which presents the signal detection efficiency as a
function of $\Delta m$. It may be noted that this efficiency reduces
drastically for $\Delta m < 5$ GeV. More or less similar results are
obtained by  L3 \cite{12},  ALEPH \cite{2} {\rm and} 
DELPHI \cite{13} colaborations. In such cases chargino pairs produced
in association with a hard photon may serve the purpose.

The weakening of chargino mass limits for small $\Delta m^{\prime}$
has also been noted by various  LEP collaborations (see, for
example, Fig 9 of \cite{2}). It is evident from this figure that
points in the $\MCH - \MSNU$ plane cannot be ruled out if $\Delta
m^{\prime}$ is small. This is true for all $\MCH$ within the kinematic
limit. If in addition the sparticle spectrum as predicted by N=1
SUGRA, with a common scalar mass $m_0$ is assumed, then even the
points in the vicinity of $\Delta m^{\prime} \approx 0$ can be
excluded by the non-observation of right sleptons at  LEP.

We, however, emphasize that for given $\MSNU$ and $\MCH$ much heavier
right-sleptons are predicted by some other models. For example one may
have additional D term contributions to scalar masses at $M_G$ or at
an intermediate scale($M_I$) \cite{14}. The right sleptons can indeed
be much heavier compared to the above SUGRA prediction for certain
choices of the D term (see e.g. eqn. of \cite{14}b). It is therefore
clear that for small $\Delta m^{\prime}$ no chargino mass can be ruled
out from direct chargino searches in a model independent way.

The regions of the parameter space corresponding to small $\Delta
m^{\prime}$ can be scanned by observing reaction (1), provided a full
calculation of the cross-section including contributions from
sneutrino exchanges are available. This will be presented in the next
section.

\section{Results and Discussions}
The matrix element squared for process (1), depicted by the Feynman
diagrams of Fig. 1, is presented in the appendix. The lighter chargino
mass and couplings are controlled by three parameters $ M_2$, $ \mu$
and tan $\beta$. The only other parameter the cross section depends on
is $\MSNU$. Thus further economy in the number of parameters cannot be
achieved by assuming either a common scalar mass at the GUT scale or a
similarly unified gaugino mass. We shall, therefore, carry out our
analysis in a model independent way. Of course model dependence (e.g.,
a specific choice of $M_1$) is unavoidable if we try to compute the
branching ratios of chargino decays. However, our main interest lies
in assesing the viability of the signal in a broad class of models
where the $\W1$ decays into soft fermions + a nearly degenerate
sparticle (either the $\LSP$ or the $\SNU$) with a large branching
ratio ($\simeq$ 1). If the fermions are soft but observable, the
signal is assumed to be background free \cite{4,5}. However, if the
fermions are too soft to be observed, we have to worry about the
irreducible SM background discussed above and introduce appropriate
cuts \cite{4,5,10} on the kinematical variables of the photon
only. Thus we estimate the observability of the signal in a broad
class of theories without introducing many specific model dependent
assumptions. 

In computing the cross section as a function of the above parameters,
we shall use the following mild cuts suggested in \cite {4,5}:
${P^{\gamma}}_T \geq 10$ GeV and $10^\circ \leq \theta_{\gamma} \leq
170^\circ $. These cuts are required to remove the radiative Bhabha
background.

The cross section in $fb$ for $\MCH$ = 80 GeV at $\sqrt{s}$ = 192 GeV as a
function of $\MSNU$ is presented in Table I (II) for tan$\beta$ = 2
(20) and several negative values of $\mu$. From these tables it is
readily seen that our results are fairly insensitive to $\mu$ and
tan$\beta$ as long as $\mu$ is significantly larger than $M_2$ which
is determined from $\MCH$. 
The last two entries in both the tables are not relevant for the signal
discussed in this paper. They are included merely to illustrate the rise
in the cross section for small $\MSNU$.

We first consider the small $\Delta m$
scenario. The results for $\MSNU \rightarrow \infty$ are obtained by
neglecting the t-channel diagrams in Fig. 1. We note that for an
integrated luminosity of 500 $pb^{-1}$ we get 25 events at LEP 192 
energies. This is in perfect agreement with \cite{4}. This number
was claimed to be adequate for discovery provided these events are
indeed background free.

Unfortunately this optimistic result is rather model dependent. As
$\MSNU$ is reduced the cross section decreases and the reduction is
quite drastic for $\MSNU \leq$ 200 GeV. For example, if $\MSNU \simeq$
100 GeV, the lighter chargino may still be the next lightest super
particle (NLSP) and also the only charged particle within the
kinematic reach of LEP 192  in many models. This is particularly so in
models in which the right- sleptons happen to be heavier than
their left counter parts \cite{14,15}. Yet we see from Table I
that if this indeed is the case, one can expect only 2-3 events 
for $\MCH$ = 80 GeVin the
single photon channel which is hardly encouraging. It is, therefore,
fair to conclude that the discovery of nearly invisible charginos 
with masses close to the kinematical limit of LEP 192  seems
to be problematic even in the single photon + missing energy channel
if the sneutrinos happen to be relatively light but still outside the reach
of the collider. For much smaller $\MCH$, the signal can be viable. For
example with $\MCH$ =65 GeV and $\MSNU$ = 100 GeV, we get 25 events.  
  
We now turn our attention to the scenario where charginos are indeed
invisible due to extremely small $\Delta m$. Here one has to take care
of the irreducible  SM background due to $ e^+e^-\longrightarrow
\nu\bar{\nu} \gamma$. Using the missing mass cut to control the
background, it was shown in \cite{4} that $\MCH\leq 65$ GeV can be
probed at  LEP 192 , if $\MSNU\rightarrow\infty$. This search limit
was obtained by the criterion $\frac{S}{\sqrt{B}} \geq 5$ where S and B
are respectively the size of the signal and the background.

We find that the cross-section for $\MCH = 65$ GeV and $\MSNU = 1000$
GeV is approximately 200 fb with the nominal cuts given above. 
The same cross-section is obtained for
$\MCH \simeq 51$ GeV and $\MSNU = 100$ GeV. A rough guess therefore
would be that only $\MCH \leq 50$ GeV can be probed if $\MSNU$ happens
to be small. This is hardly an improvement compared to the 
LEP1 limit $\MCH\geq 45$ GeV.

We note that our main conclusion remains valid even if the chargino is
mixed. In order to illustrate this we take $\MCH$ = 80 GeV, $\mu$ =
-100 GeV, $\tan\beta$ = 15. We find that $\sigma$ = 28.8, 26.79,
18.91, 8.16 fb for $\MSNU \rightarrow\infty$, 500, 200, 100 GeV,
respectively. It is amusing to note  here that the destructive
interference between the $s$ and the $t$ channel is less severe than
that for gaugino like charginos due to the significant Higgsino
component in the chargino. However, even in the $\MSNU\rightarrow\infty$ case
the cross section is small compared to the gaugino dominant scenario
due to a reduced $Z\;\CH\;\CH$ coupling. As a result even a relatively
mild destructiive interference renders the cross section unobservable
for small $\MSNU$.
    
We next consider the small $\Delta m^{\prime}$ case. A mere glance at
Table I or II would show that the number of events does
not exceed 1 for L = 500 $ {pb}^{-1}$. Thus, if charged sleptons
happens to be beyond the reach of  LEP 192 , SUSY signal would
indeed be problematic if $\Delta m^{\prime}$ is small.

However, as shown in \cite{10} a viable signal in the $\gamma + \not\!\!{E}$
channel may still be obtained through the processes\\
 $e^+e^-\rightarrow
(a)\;\SNU\BARSNU\gamma, \;(b)\;\LSP\N2\gamma, \; and\; (c)\;\LSP\LSP\gamma,$\\
 where both $\SNU$ and $\N2$ decay invisibly. If in addition $\CH$'s decay
invisibly due to very small $\Delta m^\prime$ the statistical significance
of the signal will be further enhanced due to (d) the process of Eq.1.
As an example we present the cross section of process (a) to (d) at
$\sqrt s$ = 192 for $\MSNU$ = 63.5, $\MCH$ = 65.0, $\mu$ = -500 (all in GeV)
 and $tan\beta$ = 2, we obtain in ($pb$). $\sigma_a$ = 0.063, $\sigma_b$ = 0.03,
$\sigma_c$ = 0.035, $\sigma_d$ = 0.037.

In completing the above cross sections we have used the cuts of \cite{16}
 ${p_\gamma}^T \geq$ 6.17, 6.175 $\leq E_\gamma\leq$ 47.5,
  $18^o\leq \theta_\gamma\leq 162^o$. The SM
background from $e^+e^-\rightarrow \nu\bar{\nu}\gamma$ with these cuts is
 0.95 $pb$. For an integrated luminosity of 500 ${pb}^{-1}$, the statistical
significance $\frac{S}{\sqrt{B}}$ (as defined above) is 4. Thus 
$\MCH\approx\MSNU\leq$ 65 GeV is expected to yield an observable signal
at LEP 192 . As shown in \cite{10} much larger values of $\MCH(\simeq\MSNU)$
 can be probed at NLC energies.

For nearly invisible charginos on the other hand the results of the last
paragraph predict
18 (or more) background free events for $\MCH\leq$ 65 GeV. It is therefore
appears that the search limits for both invisible and nearly invisible
charginos at LEP 192 are similar. 

We now turn our attention to  NLC energies, $\sqrt{s}$ = 500
GeV. It was shown in \cite{4} that if charginos decay nearly invisibly
to a background free final state then $\geq 50$ events can be expected for
$\MCH\leq$ 240 GeV.  This is encouraging since it suggests that the
search can be extended almost up to the kinematical limit. Using our
formula we get a cross-section of 1 fb for $\MCH$ = 240 GeV, $\MSNU 
\rightarrow\infty$, $\mu$ = -500 and $\tan\beta$ = 2 in agreement with
\cite{4}. If $\MSNU$ is now lowered keeping the other parameters
fixed, then for $\MSNU$ = 200, the cross-section drops to 0.01 fb
which predicts no event for an integrated luminosity of 50
${fb}^{-1}$.

Although we have presented our results for negative $\mu$ only, we have 
checked that our main conclusion remain valid for positive $\mu$ as long
as the chargino is in the deep gaugino or in the mixed region.

\section{Summary and conclusions}

The main result of this paper is the computation of the cross section
for the process $e^+e^- \rightarrow \widetilde{\chi}^+
\widetilde{\chi}^- \gamma$ in a model independent way. In previous
calculations \cite{4,5} this was computed in the special case $\MSNU
\rightarrow \infty$.

As pointed out in \cite{4,5} the above process is very important for
the discovery of "invisible" or "nearly invisible" charginos which
decay dominantly into soft leptons (or hadrons) + $\not\!\!{E}$ due to
their degeneracy with the LSP. We find that if $\MSNU$ is relatively
small but still outside the kinematical reach of the collider, the
cross section is strongly suppressed to the level of unobservability,
due to destructive interferences in the matrix element squared. This
is indeed serious, since such a chargino may happen to be the only
 visible sparticle within the striking range of
 LEP. For example, the optimistic search limit $\MCH\leq$ 65 GeV
energies for invisibly decaying charginos obtained in \cite{4,5}
for $\MSNU\rightarrow\infty$, is reduced to $\MCH\leq$ 50 GeV for
$\MSNU\simeq$ = 100 GeV.  
New strategies for hunting down such charginos should be
evolved. At NLC energies the optimistic search limits for these
charginos obtained in \cite{4,5} may be reduced considerably if
sneutrinos are relatively light.

We have also noted that the chargino may decay invisibly or nearly
invisibly due to its degeneracy or near-degeneracy with the
sneutrinos. In this case the  signal from charginos alone is necessarily
unobservable. However, as pointed out in \cite{10}, SUSY signals may still
be searched in the $\gamma + \not\!\!{E}$ channel via 
$e^+e^-\rightarrow \SNU\BARSNU\gamma$, $\LSP\N2\gamma$, $\LSP\LSP\gamma$,
  where both $\SNU$'s and $\N2$ decay invisbly.
 Unfortunately at  LEP 192  energies the
signal is marginal even for favourable choices of $\MSNU$. Additional
contributions from  
$e^+e^- \rightarrow \tilde{\chi}^+ \tilde{\chi}^- \gamma$ may enhance the
statistical significance of the signal if the charginos decay invisibly.
A viable signal can be obtained at LEP 192  energies if $\MCH\simeq\MSNU\leq$
 65 GeV.   At NLC, however resonably a larger region of the parameter space.

\vskip 10pt
\noindent
{\bf Acknowledgements:} The authors wish to thank M. Drees for a critical
reading of the manuscript and many valuable comments. The work of AD was
supported by BRNS, grant no. 37/4/97 -G and DST,  grant no.
 SP/S2/K -01/97, Govt. of India. The work of SM was supported by UGC of India.

\newpage
\noindent
FIGURE CAPTIONS
\vskip 2pt
\noindent
FIGURE 1 \hskip 5pt\rm The Feynman diagrams corresponding to the process   
$e^+e^- \rightarrow \tilde{\chi}^+ \tilde{\chi}^- \gamma$.
\vskip 20pt
\noindent
TABLE CAPTIONS
\vskip 2pt
\noindent
TABLE-I: \hskip 5pt\rm The signal cross section in $fb$ as a function of
$\MSNU$ for $tan\beta$ = 2 (see text for further details).
\vskip 10pt
\noindent
TABLE-II: \hskip 5pt\rm Same as TABLE-I but for $tan\beta$ = 20.
\newpage
\begin{center}
{\bf TABLE-I}
\end{center}
\begin{center}
\begin{tabular}{||c|c|c|c|c||}
\hline
\hline
&$\mu = -200$&$\mu=-300$&$\mu=-400$&$\mu=-500$\\
\hline
&$M_2 =64.0$&$M_2=67.33$&$M_2=70.0$&$M_2=71.66$\\
\hline
$\MSNU$ & $\sigma_{total}$ & $\sigma_{total}$ & 
$\sigma_{total}$ & $\sigma_{total}$ \\
\hline
\hline
$\infty$ & 49.97 & 52.64 & 50.81 & 51.81 \\
1000 & 47.81 & 50.48 & 49.12 & 49.39 \\
500 & 43.84 & 46.35 & 45.14  & 45.44 \\
200 & 25.28 & 27.17 &26.59 & 26.79 \\
100 & 5.31 & 6.27 & 6.21 & 6.33 \\
90 &  3.62 & 4.44 & 4.41 & 4.52 \\
80 &  2.41 & 3.06 & 3.05 & 3.15\\
70 &  1.89 &2.36 &  2.36 & 2.44 \\
60 &  2.29 &2.56 &  2.51 & 2.56 \\
\hline
\hline
\end{tabular}
\end{center}

\vskip 8pt
\begin{center}
{\bf TABLE-II}
\end{center}
\begin{center}
\begin{tabular}{||c|c|c|c|c||}
\hline
\hline
&$\mu = -200$&$\mu=-300$&$\mu=-400$&$\mu=-500$\\
\hline
&$M_2 =90.66$&$M_2=84.0$&$M_2=81.66$&$M_2=81.0$\\
\hline
$\MSNU$ & $\sigma_{total}$ & $\sigma_{total}$ & 
$\sigma_{total}$ & $\sigma_{total}$ \\
\hline
\hline
$\infty$ & 46.34 & 49.23 & 53.08 & 51.96 \\
1000 & 44.41 & 47.57 & 50.74 & 49.94 \\
500 & 40.74 & 43.64 & 46.58 & 45.89 \\
200 & 23.34 & 25.30 &27.29 & 26.95 \\
100 & 4.63 & 5.49 & 6.29 & 6.27 \\
90 &  3.05 & 3.78 & 4.45 & 4.46 \\
80 &  1.91 & 2.54 & 3.08 & 3.10\\
70 &  1.45 &1.97 &  2.40 & 2.41 \\
60 &  1.85 &2.30 &  2.63 & 2.61 \\
\hline
\hline
\end{tabular}
\end{center}
\newpage
\def \CH{\widetilde{\chi}^{\pm}}
\def \MCH{m_{\widetilde{\chi}^{\pm}}}
\def \MCHSQ{m_{\widetilde{\chi}^{\pm}}^2}
\def \SNU{\tilde{\nu}}
\def \MSNU{m_{\tilde{\nu}}}
\def \MSNUSQ{m_{\tilde{\nu}}^2}
\def \MZSQ{M_Z^2}
\def \GVSQ{g_V^2}
\def \GASQ{g_A^2}
\def \ASQ{a^2}
\def \BSQ{b^2}
\def \GZ{{\Gamma}_Z}
\def \GW{{\Gamma}_W}
\renewcommand{\P}[2]{P_{#1#2}}

\vskip 8pt
\begin{flushleft}
{\LARGE \bf Appendix}
\end{flushleft}
\vskip 20pt
\noindent
In this appendix we systematically present the relevant formulae
for calculating the cross sections of different processes.
Throughout this paper we use the following Standard Model
Parameters :\\

            $ \alpha = 1/128.8, \;
              G_F   = 1.16637 \, \times 10^{-5}, \;
              M_Z   = 91.187 \, GeV, \;
              M_W   = 80.33  \, GeV, \;
              {\Gamma}_Z   =  2.49 \, GeV, \;
              T_3^e  = -0.5,        \;  
              Q_e   = -1, \; 
              S_W^2  = sin^2{\theta_{W}} = 0.232, \;
              C_W = cos{\theta_{W}}, \;
              g_V   =  0.5 T_3^e -  Q_e S_W^2, \;
              g_A   = -0.5 T_3^e,\;
              S_{avg} = 1/4$(spin averaging over the
                                 initial spin configuration)

\vskip 25pt
\begin{flushleft}
{\Large \bf The process $e^+ \, e^- \longrightarrow 
\widetilde{\chi}^+ \widetilde{\chi}^- \, + \, \gamma$.}
\end{flushleft}
\vskip 10pt
\noindent
We label the particles by the following indices:
 $e^+ \Rightarrow 1 , \;
 e^- \Rightarrow 2 , \;
 \widetilde{\chi}^- \Rightarrow 3 , \;
 \widetilde{\chi}^+ \Rightarrow 4 , \;
 \gamma \Rightarrow 5$ .
\noindent
We have used the following abbreviations :
              $ P_{ij} = p_i.p_j, \;
\epsilon{(ijkl)}= \epsilon_{\alpha \beta \gamma \delta}
p_i^{\alpha} p_j^{\beta} p_k^{\gamma} p_l^{\delta} $, where
$p_i$ is the momemtum of the $i-th$ particle.\

\vskip 3pt
\noindent
In the following ${\bf T_{ij}} $ = $A_i A_j^ \dagger +$ H.C.,
where $A_i$ is the amplitude of the $i-th$ Feynman diagram.\\
 $ov_1 = 32\pi\alpha G_F^2 M_W^4/C_W^4$, \hskip 6pt
 $ov_2 = 32\sqrt 2 {\pi}^2 {\alpha}^2 G_F M_W^2/C_W^2$,\hskip 6pt 
 $ov_3 = (4\pi\alpha)^3$\\
 $ov_4 = 2\pi\alpha G_F^2 M_W^4 V_{21}^4$,\hskip 6pt
 $ov_5 = 4\pi\alpha G_F^2 M_W^4 V_{21}^2/C_W^2$,\hskip 6pt
 $ov_6 = 4\sqrt 2 (\pi\alpha)^2 M_W^2 G_F V_{21}^2$\\
$ O^L$= $-V_{21}^2 -0.5V_{22}^2 + S_W^2$,\hskip 6pt 
$O^R$ = $-U_{21}^2 -0.5U_{22}^2 + S_W^2$\\ 
a = $ O^R + O^L$,\hskip 10pt
b = $ O^R - O^L$,\hskip 10pt
$S_1= 2P_{12} - M_Z^2.$\\
$S_2= 2P_{34} - M_Z^2 + 2\MCHSQ $,\hskip 6pt
$Z_1=1/(s_1^2 + (M_Z\GZ)^2)$\\
$Z_2=1/(s_2^2 + (M_Z\GZ)^2)$,\hskip 6pt
$Z_3=1/(s_2 + M_Z^2)$\\
$sn_1=\MCHSQ - \MSNUSQ - 2P_{14}$,\hskip 6pt
$sn_2=\MCHSQ - \MSNUSQ - 2P_{23}$\\
$C_1= ab(\GASQ + \GVSQ) + g_Ag_V(\ASQ + \BSQ)$,\hskip 6pt
$C_2=ab(\GASQ + \GVSQ) - g_Ag_V(\ASQ + \BSQ)$
where $ V_{ij}$ and  $ U_{ij}$ are the elements of unitary matrices {\bf
U} and {\bf V} which diagonalise the chargino mass matrix.
\newpage
The relevant matrix element squared can be computed from the
following formulae:\\
(1)First we consider the $s$-channel Z-exchange diagrams (1-4) in Fig. 1. 
\begin{flushleft}
$ {\bf T_{11}}= -32.Z_2R_{11}/P_{15}$
\vskip 5pt
$R_{11}= (\GVSQ + \GASQ) \{ -(\ASQ + \BSQ)(P_{23}P_{45} +P_{24}P_{35})
 + (\BSQ - \ASQ)\MCHSQ P_{25}\}  + 4.g_A g_V ab (P_{24} P_{35}
 - P_{23} P_{45})$
\end{flushleft}
\begin{flushleft}
$ {\bf T_{22}}= -32.Z_2R_{22}/P_{25}$
\vskip 5pt
$R_{22}=(\GVSQ + \GASQ) \{ -(\ASQ + \BSQ)(P_{13}P_{45}
 +P_{14}P_{35})
 + (\BSQ - \ASQ)\MCHSQ P_{15}\}  + 4.g_A g_V ab (P_{14} P_{35} +
 P_{13} P_{45})$
\end{flushleft}
\begin{flushleft}
$ {\bf T_{33}}= -32.Z_1R_{33}/P_{35}^2$
\vskip 5pt
$ R_{33}=(\GVSQ + \GASQ)[(\ASQ + \BSQ)\{(P_{14}(P_{23} + P_{25})+
 P_{24}(P_{13} + P_{15}))\MCHSQ - P_{35}(P_{15}P_{24} + P_{14}P_{25})\} +
           (\ASQ - \BSQ)(P_{35} + \MCHSQ)P_{12}\MCHSQ] + 4.g_Ag_Vab\{
           (P_{35} - \MCHSQ)(P_{15}P_{24} -P_{14}P_{25}) + \MCHSQ(P_{14}P_{23}
      - P_{13}P_{24})\}$
\end{flushleft}
\vskip 5pt
\begin{flushleft}
$ {\bf T_{44}}= -32.Z_1R_{44}/P_{45}^2$
\vskip 5pt
$R_{44}= (\GVSQ + \GASQ)[(\ASQ + \BSQ)\{(P_{13}(P_{24} + P_{25})+
 P_{23}(P_{14} + P_{15}))\MCHSQ - P_{45}(P_{15}P_{23} + P_{13}P_{25})\} +
 (\ASQ - \BSQ)(P_{45} + \MCHSQ)P_{12}\MCHSQ] + 4.g_Ag_Vab\{
 (P_{45} - \MCHSQ)(P_{13}P_{25} -P_{15}P_{23}) + \MCHSQ(P_{14}P_{23} -
           P_{13}P_{24})\}$
\end{flushleft}
\vskip 5pt

\begin{flushleft}
$ {\bf T_{12}}= 32.Z_2R_{12}/P_{15}/P_{25}$
\vskip 5pt
$R_{12}=(\GASQ + \GVSQ)[(\ASQ + \BSQ)\{X_{12} +Y_{12} +2\P15\P23\P24 +
 2\P13\P14\P25\} + 2.(\ASQ - \BSQ)\P12(\P12 -\P15 - \P25)\MCHSQ]
 + 4.g_Ag_Vab\{X_{12} - Y_{12}\}$\\
$X_{12}= \P12\P14(2\P23 - \P35) - \P14\P23(\P15 + \P25) -\P12\P23\P45$
\vskip2 pt
$Y_{12}= X_{12}(3\longleftrightarrow 4)$
\end{flushleft}
\begin{flushleft}
${\bf T_{13}}=-32.Z_1 Z_2[ \{S_1S_2 + (M_Z \GZ)^2\}R_{13}
  - M_Z\GZ(S_1 - S_2)I_{13}] /\P15/\P35$
\vskip 2pt
$R_{13}=(\GASQ + \GVSQ)[(\ASQ + \BSQ)(X_{13} + Y_{13})
 + \MCHSQ(\ASQ - \BSQ)M_{13} + \MCHSQ(\ASQ + \BSQ)\P15\P24] +
 4.g_Ag_Vab(X_{13} - Y_{13}  -  \P15\P24\MCHSQ)$
$X_{13}=\P14\P23(2\P13 + P15) - \P14\P35(\P12 + \P23) + \P13\P14\P25
+\P15\P23\P34 - \P13\P23\P45$
\vskip 2pt
$Y_{13}= 2\P24(\P13 + \P15)(\P13 - \P35)$
\vskip 1pt
$M_{13}=2.\P12\P13 + \P12\P15 + \P15\P23  - \P13\P25
 - \P15\P25 - \P12\P35$
\vskip 2pt
$I_{13}=C_1[(\P14 + \P34)\epsilon{(5123)} + (\P13 - \MCHSQ)
\epsilon{(5124)} - (\P12 + \P23)\epsilon{(5134)} -
\P13\epsilon{(5234)} + (\P35 - \P15)\epsilon{(1234)}]
 + 2.g_Ag_V(\ASQ - \BSQ)\MCHSQ\epsilon{(5123)}$
\end{flushleft}
\begin{flushleft}
${\bf T_{14}}=32.Z_1 Z_2[ \{S_1S_2 + (M_Z \GZ)^2\}R_{14}
  - M_Z\GZ(S_1 - S_2)I_{14}] /\P15/\P45$

$R_{14}=(\GASQ + \GVSQ)[(\ASQ + \BSQ)\{Y_{13}(3\leftrightarrow 4)
 +X_{13}(3\leftrightarrow 4)\}
+\MCHSQ(\ASQ - \BSQ)M_{13}(3\leftrightarrow 4) + \MCHSQ(\ASQ + \BSQ)\P15\P23]
 + 4g_Ag_Vab\{Y_{13}(3\leftrightarrow 4) - X_{13}(3\leftrightarrow 4) +
 \P15\P23\MCHSQ\}$
\vskip 2pt
$I_{14}=C_2[(\P14 - \MCHSQ)\epsilon{(5123)} +
(\P13 + \P34)\epsilon{(5124)}+ (\P12 + \P24)\epsilon{(5134)}+
 \P14\epsilon{(5234)} +(\P15 - \P45)\epsilon{(1234)}] 
+ 2g_Ag_V\MCHSQ(\ASQ - \BSQ)\epsilon{(5124)}$
\end{flushleft}
\begin{flushleft}
${\bf T_{23}}=32.Z_1 Z_2[ \{S_1S_2 + (M_Z \GZ)^2\}R_{23}
  - M_Z\GZ(S_1 - S_2)I_{23}] /\P25/\P35$
\vskip 2pt
$R_{23}=(\GASQ + \GVSQ)[(\ASQ + \BSQ)\{Y_{13}(1\longleftrightarrow 2)
+ X_{13}(1\longleftrightarrow 2)\} +\MCHSQ(\ASQ - \BSQ)(2\P12\P23 - \P15\P23 +
 \P12\P25 + \P13\P25  - \P15\P25 - \P12\P35) + \MCHSQ(\ASQ + \BSQ)\P14\P25]
+ 4g_Ag_Vab\{Y_{13}(1\longleftrightarrow 2)
- X_{13}(1\longleftrightarrow 2) + \P14\P25\MCHSQ\}$
\vskip 2pt
$I_{23}= g_Ag_V(\ASQ + \BSQ)\{(2\P24 - \P45)\epsilon{(5123)}
 + \P35\epsilon{(5124)} + \P25(-\epsilon{(5134)} + \epsilon{(1234)}) +
 (2\P13 + \P15)\epsilon{(5234)}\}
 + ab(\GASQ + \GVSQ)\{ - (\P24 + \P34)\epsilon{(5123)} -
 \P23\epsilon{(5134)} - (\P12 + \P13)\epsilon{(5234)} -
 \P35\epsilon{(1234)}\} + 2g_Ag_V\MCHSQ(\ASQ - \BSQ)\epsilon{(5123)}$
\end{flushleft}
\begin{flushleft}
${\bf T_{24}}=-32.Z_1 Z_2[ \{S_1S_2 + (M_Z \GZ)^2\}R_{24}
  - M_Z\GZ(S_1 - S_2)I_{24}] /\P25/\P45$
\vskip 2pt
$R_{24}=(\GASQ + \GVSQ)[(\ASQ + \BSQ)\{X_{13}(1\leftrightarrow 2 \;\&\; 3
\leftrightarrow 4) +Y_{13}(1\leftrightarrow 2 \;\&\; 3\leftrightarrow 4) + 
\MCHSQ(\ASQ + \BSQ)(2\P12\P24 - \P15\P24 + \P12\P25  +
\P14\P25 - \P15\P25 - \P12\P45) + \MCHSQ(\ASQ + \BSQ)\P13\P25] +
4g_Ag_Vab\{X_{13}(1\leftrightarrow 2 \;\&\; 3\leftrightarrow 4) - Y_{13}
(1\leftrightarrow 2 \;\&\; 3\leftrightarrow 4) - \P13\P25\MCHSQ\}$
\vskip 2pt
$I_{24}= g_Ag_V(\ASQ + \BSQ)\{\P45\epsilon{(5123)} +
(2\P23 - \P35)\epsilon{(5124)} + \P25\epsilon{(5134)}
- (2\P14 + \P15)\epsilon{(5234)}\} +ab(\GASQ +\GVSQ)\{\P24(\epsilon{(5123)}
 - \epsilon{(5134)}) + (\P23 + \P34)\epsilon{(5124)}
 - (\P14 + \P12)\epsilon{(5234)} +(\P25 - \P45)\epsilon{(1234)} -
 \MCHSQ\epsilon{(5123)}\}$
\end{flushleft}
\begin{flushleft}
${\bf T_{34}}=32.Z_1 R_{34}/P_{35}/P_{45}$
\vskip 2pt
$R_{34}=(\GASQ + \GVSQ)(\ASQ + \BSQ)\{X_{34} + Y_{34} - 2\P14\P24\P35
- 2\P13\P23\P45\} +(\GASQ + \GVSQ)(\ASQ - \BSQ)\MCHSQ(- 2\P15\P25 +
 2\P12\P34 + \P12\P35 + \P12\P45) +
 4g_Ag_Vab(X_{34} - Y_{34})$
\vskip 2pt
$X_{34}=(2\P14 + \P15)\P23\P34 + \P14\P25\P34 + \P14\P23(\P35 +\P45)$
$Y_{34}=(2\P13 + \P15)\P24\P34 + \P13\P25\P34 + \P13\P24\P35 + \P13\P24\P45$ 
\end{flushleft}
ZZ = $ov_1({\bf T_{11}} + {\bf T_{22}} + {\bf T_{33}} + {\bf T_{44}} +
{\bf  T_{12}} + {\bf T_{13}} + {\bf T_{14}}
+ {\bf T_{23}} + {\bf T_{24}} + {\bf T_{34}})$\\
\noindent
(ZZ :represents the total Z-exchange matrix element squard)
\vskip 5pt
\noindent
(2) Here we consider $s$-channel $\gamma$ exchange diagrams (5-8 in Fig. 1).

\begin{flushleft}
${\bf T_{55}}= 32Z_3^2(\P24\P35 + \P23\P45 + \P25\MCHSQ)/\P15$
\end{flushleft}
\begin{flushleft}
${\bf T_{56}}=-32Z_3^2\Big [ X_{16} + X_{16}(3\leftrightarrow 4)
- 2\P15\P23\P24 - 2\P13\P14\P25 +2\P12(-P12 +
 \P15 +\P25)\MCHSQ\Big ]/\P15/\P25$
\end{flushleft}
\begin{flushleft}
${\bf T_{57}}=-16Z_3\Big [ X_{13} + Y_{13} + \MCHSQ(M_{13} +
 \P15\P24)\Big ]/\P12/\P15/\P35$
\end{flushleft}
\begin{flushleft}
${\bf T_{58}}=16Z_3\Big [ X_{13}(3\leftrightarrow 4) +
 Y_{13}(3\leftrightarrow 4) + \MCHSQ\{M_{13}(3\leftrightarrow 4) +
 \P15\P23\}\Big ]/\P12/\P15/\P45$
\end{flushleft}
\begin{flushleft}
${\bf T_{66}}=32Z_3^2(\P14\P35 + \P13\P45 + \P15\MCHSQ)/P25$
\end{flushleft}
\begin{flushleft}
${\bf T_{67}}=16Z_3\Big [ X_{13}(1\leftrightarrow 2) +
 Y_{13}(1\leftrightarrow 2) + \MCHSQ(M_{13}(1\leftrightarrow 2) +
 \P14\P25)\Big ]/\P12/\P25/\P35$
\end{flushleft}
\begin{flushleft}
${\bf T_{68}}=-16Z_3\Big [ X_{13}(1\leftrightarrow 2 \;\&\; 3\leftrightarrow 4) +
 Y_{13}(1\leftrightarrow 2 \;\&\; 3\leftrightarrow 4) +
 \MCHSQ\{M_{13}(1\leftrightarrow 2 \;\&\; 3\leftrightarrow 4) +
 \P13\P25\}\Big ]/\P12/\P25/\P45$
\end{flushleft}
\begin{flushleft}
${\bf T_{77}}= -8\Big [-\P15\P24\P35 - \P14\P25\P35
+ (\P14\P23 + \P13\P24 + \P15\P24   + \P14\P25 + \P12\P35
+ \P12\MCHSQ)\MCHSQ) \Big ]/\P12^2/\P35^2$
\end{flushleft}
\begin{flushleft}
${\bf T_{78}}=8\Big [X_{34} + Y_{34} - 2\P14\P24\P35 - 2\P13\P23\P45 +
(- 2\P15\P25 + 2\P12\P34 + \P12\P35 + \P12\P45)\MCHSQ\Big ]/\P12^2/\P35/\P45$
\end{flushleft}
\begin{flushleft}
${\bf T_{88}}= -8\Big [-\P15\P23\P45 - \P13\P25\P45 + (\P14\P23 + \P15\P23 +
 \P13\P24 + \P13\P25 + \P12\P45 + \P12\MCHSQ)\MCHSQ\Big ]/\P12^2/\P45^2$
\end{flushleft}
\begin{flushleft}
GG = $ov_3({\bf T_{55}} + {\bf T_{56}} + {\bf T_{57}} + {\bf T_{58}} +
 {\bf T_{66}} + {\bf T_{67}}
 + {\bf T_{68}} +{\bf T_{77}} + {\bf T_{78}} + {\bf T_{88}})$
\end{flushleft}
\noindent
(GG: represents the total $\gamma$-exhange matrix element squard)
\vskip 5pt
\noindent
(3)Interference between Z-exchange and $\gamma$-exchange  
diagrams is considered here.

\begin{flushleft}
${\bf T_{15}}=64.S_2Z_2Z_3 R_{15}/P_{15}$
\vskip 1pt
$R_{15}= (\P24\P35 + \P23\P45
 + \P25\MCHSQ)g_Va +(-\P24\P35 + \P23\P45)g_Ab$
\end{flushleft}

\begin{flushleft}
${\bf T_{16}}= -32Z_2Z_3\{S_2R_{16} - M_Z\GZ I_{16}\}/\P15/\P25$
\vskip 2pt
$R_{16}= \{X_{16} + X_{16}(3\leftrightarrow 4) -
 2\P15\P23\P24 - 2\P13\P14\P25\}ag_V + 2(-\P12 + \P15 + \P25)\P12ag_V\MCHSQ +
 \{X_{16} - X_{16}(3\leftrightarrow 4)\}bg_A$
\vskip 2pt
$X_{16}=\P14\P23(-2\P12 +\P15 +\P25) + \P12\P14\P35 + \P12\P23\P45$
\vskip 1pt
$I_{16}= \{(- \P14 + \P24)\epsilon{(5123)} +
(- P13 + P23)\epsilon{(5124)}\}ag_A + \{\P12\epsilon{(5134)}
 - \P12\epsilon{(5234)} + (\P15 + \P25)\epsilon{(1234)}\}bg_V $
\end{flushleft}
\begin{flushleft}
${\bf T_{17}}= -16Z_2\{S_2R_{17} - M_Z\GZ I_{17}\}/\P15/\P12/\P35$
\vskip 2pt
$R_{17}= (X_{13} + Y_{13})ag_V
 + (2\P12\P13 + \P12\P15 + \P15\P23 + \P15\P24 - 
\P13\P25 - \P15\P25 - \P12\P35)ag_V\MCHSQ + (X_{13} - Y_{13})bg_A
 + (\P12\P15 - \P15\P24 - \P15\P25)bg_A\MCHSQ$  
\vskip 2pt
$I_{17}=0.5\{(\P14ag_A + \P34ag_A + 2ag_A\MCHSQ
 + \P14bg_V + \P34bg_V)\epsilon{(5123)} + (\P13ag_A -a g_A\MCHSQ + \P13b g_V  
 -b g_V\MCHSQ)\epsilon{(5124)} + (- \P12ag_A - \P23a g_A - \P12b g_V  
 - \P23b g_V)\epsilon{(5134)} + (-\P13a g_A - \P13b g_V)\epsilon{(5234)}
 +(- \P15a g_A + \P35a g_A - \P15b g_V +  \P35b g_V)\epsilon{(1234)}\}$
\end{flushleft}
\begin{flushleft}
${\bf T_{18}}= 16Z_2\{S_2R_{18} - M_Z\GZ I_{18}\}/\P12/\P15/\P45$
\vskip 1pt
$R_{18}= \{Y_{13}(3\leftrightarrow 4) + X_{13}(3\leftrightarrow 4)\}ag_V
 +(2\P12\P14 + \P12\P15 + \P15\P23 + \P15\P24 - 
 \P14\P25 - \P15\P25 - \P12\P45)ag_V\MCHSQ + 
\{Y_{13}(3\leftrightarrow 4) - X_{13}(3\leftrightarrow 4)\}bg_A
 -(\P12\P15 + \P15\P23 + \P15\P25)bg_A\MCHSQ$  
\vskip 2pt
$I_{18}=0.5\{ (\P14ag_A - ag_A\MCHSQ - \P14bg_V
 + bg_V\MCHSQ)\epsilon{(5123)} + (\P13ag_A + \P34ag_A + 2ag_A\MCHSQ - \P13bg_V 
 - \P34bg_V)\epsilon{(5124)} +(\P12ag_A + \P24ag_A - \P12bg_V  
 - \P24bg_V)\epsilon{(5134)} +(\P14ag_A - \P14bg_V)\epsilon{(5234)}
 + (\P15ag_A - \P45ag_A - \P15bg_V  + \P45bg_V)\epsilon{(1234) }$
\end{flushleft}
\begin{flushleft}
${\bf T_{25}}= 32Z_2Z_3\{S_2R_{25} - M_Z\GZ I_{25}\}/\P15/\P25$
\vskip 2pt
$R_{25}= \{X_{25} + X_{25}(1\leftrightarrow 2) + 2\P15\P23\P24 +
 2\P13\P14\P25\}ag_V + 2\P12(\P12 - \P15 - \P25)ag_V\MCHSQ + 
\{X_{25} - X_{25}(1\leftrightarrow 2)\}bg_A$
\vskip 2pt
$X_{25}= \P14\P23(2\P12 - \P15 - \P25) - \P12(\P14\P35 + \P23\P45)$
\vskip 1pt
$I_{25}=ag_A\{(\P24 - \P14)\epsilon{(5123)} +
 (\P23 - \P13)\epsilon{(5124)}\}  
 + bg_V\{\P12\epsilon{(5134)} - \P12\epsilon{(5234)} +  
(\P15 + \P25)\epsilon{(1234)}\}$ 
\end{flushleft}
\begin{flushleft}
${\bf T_{26}}=64Z_2Z_3S_2\Big [ 
ag_V(\P14\P35 + \P13\P45
+ \P15\MCHSQ) + bg_A(\P14\P35 - \P13\P45)\Big ]/P25$
\end{flushleft}
\begin{flushleft}
${\bf T_{27}}= 16Z_2\{S_2R_{27} - M_Z\GZ I_{27}\}/\P12/\P25/\P35$
\vskip 2pt
$R_{27}= \{Y_{13}(1\leftrightarrow 2)+ X_{13}(1\leftrightarrow 2)\}ag_V
+ (2\P12\P23 -  \P15\P23 + \P12\P25 + \P13\P25 + 
 \P14\P25 - \P15\P25 - \P12\P35)ag_V\MCHSQ + \{Y_{13}(1\leftrightarrow 2)
- X_{13}(1\leftrightarrow 2)\}bg_A + (- \P12\P25 +  \P14\P25 +
 \P15\P25)bg_A\MCHSQ)$
\vskip 2pt
$I_{27}=0.5\Big [\{ag_A(2\P24 - \P45 +  2\MCHSQ) - bg_V(\P24  
 - \P34)\}\epsilon{(5123)} + (\P35ag_A  - \P23bg_V +
 bg_V\MCHSQ)\epsilon{(5124)} -(\P25ag_A + \P23bg_V)\epsilon{(5134)} + 
\{ag_A(2\P13 + \P15)  - bg_V(\P12 + \P13)\}\epsilon{(5234)} + 
bg_V(P25 - P35)\epsilon{(1234)}\Big ]$
\end{flushleft}
\begin{flushleft}
${\bf T_{28}}= - 16Z_2\{S_2R_{28} - M_Z\GZ I_{28}\}/\P12/\P25/\P45$
\vskip 2pt
$R_{28}= \{X_{13}(1\leftrightarrow 2 \;\&\; 3\leftrightarrow 4)+
 Y_{13}(1\leftrightarrow 2 \;\&\; 3\leftrightarrow 4)\}ag_V +
\{(2\P12 - \P15)\P24 + (\P12 + \P13 + 
\P14 - \P15)\P25 - \P12\P45\}ag_V\MCHSQ + 
\{X_{13}(1\leftrightarrow 2 \;\&\; 3\leftrightarrow 4)-
 Y_{13}(1\leftrightarrow 2 \;\&\; 3\leftrightarrow 4)\}bg_A +
 \P25(\P12 - \P13 - \P15)bg_A\MCHSQ $
\vskip 2pt
$I_{28}=0.5\Big [ (\P45ag_A + \P24bg_V - bg_V\MCHSQ)\epsilon{(5123)} +
\{(2\P23 - \P35 + 2\MCHSQ)ag_A + (\P23 + P34)bg_V\}\epsilon{(5124)} + 
(\P25ag_A - \P24bg_V)\epsilon{(5134)}   -\{ (2\P14 + P15)ag_A + 
(\P12 + P14)bg_V\}\epsilon{(5234)} + (\P25 - \P45)bg_V\epsilon{(1234)}\Big ]$
\end{flushleft}
\begin{flushleft}
${\bf T_{35}}= -32Z_1Z_3\{S_1R_{35} - M_Z\GZ I_{35}\}/\P15/\P35$
\vskip 2pt
$R_{35}= (X_{13} + Y_{13})ag_V
+ \{\P12(2\P13 + \P15 - \P35) + \P15(\P23 +
 \P24 - \P25) - \P13\P25\}ag_V\MCHSQ + (X_{13} - Y_{13})bg_A
+ (- \P12\P15 - \P15\P24 + \P15\P25)bg_A\MCHSQ $ 
\vskip 2pt
$I_{35}=0.5\Big [\{ag_A(-\P14 - P34 - 2\MCHSQ) -
 (2\P14 - P45)bg_V\}\epsilon{(5123)} + (-\P13ag_A + 
ag_A\MCHSQ - \P35bg_V)\epsilon{(5124)} + \{ ag_A(\P12
+ \P23) + bg_V(2\P23 + \P25)\}\epsilon{(5134)} + 
(\P13ag_A - \P15bg_V)\epsilon{(5234)} + 
ag_A(\P15 - P35)\epsilon{(1234)}\Big ] $
\end{flushleft}
\begin{flushleft}
${\bf T_{36}}= 32Z_1Z_3\{S_1R_{36} - M_Z\GZ I_{36}\}/\P25/\P35$
\vskip 2pt
$R_{36}= \{Y_{13}(1\leftrightarrow 2)+
 X_{13}(1\leftrightarrow 2 )\}ag_V +
 + (2\P12\P23 - \P15\P23 + \P12\P25 +
 \P13\P25 + \P14\P25 - \P15\P25 - \P12\P35)ag_V\MCHSQ + 
\{Y_{13}(1\leftrightarrow 2) - X_{13}(1\leftrightarrow 2 )\}bg_A +
 + (\P12 + \P14 - \P15)\P25bg_A\MCHSQ $
\vskip 2pt
$I_{36}=0.5\Big [\{ag_A(-P24 - P34 - 2\MCHSQ) +
 (2\P24 - \P45)bg_V\}\epsilon{(5123)} +\{(-\P23 + \MCHSQ)ag_A +
 \P35bg_V\}\epsilon{(5124)} + (-\P23ag_A - \P25bg_V)\epsilon{(5134)} +
 \{ag_A(-P12 - P13) + bg_V(2\P13 + \P15)\}\epsilon{(5234)} + 
 (\P25 - \P35)ag_A\epsilon{(1234)}\Big ]$
\end{flushleft}
\begin{flushleft}
${\bf T_{37}}= -32Z_1S_1R_{37} /\P12/\P35^2$
\vskip 2pt
$R_{37}= (ag_V + bg_A)\P14\{-\P25\P35 + \MCHSQ(\P23 + \P25)\} +(ag_V -
 bg_A)\P24\{-\P15\P35 + \MCHSQ(\P13 + \P15)\} + ag_V\MCHSQ(\P12\P35 + \MCHSQ)$
\end{flushleft}
\begin{flushleft}
${\bf T_{38}}= -16Z_1\{S_1R_{38} - M_Z\GZ I_{38}\}/\P12/\P35/\P45$
\vskip 2pt
$R_{38}=(X_{34} + Y_{34} - 2\P14\P24\P35 - 2\P13\P23\P45)ag_V
 + (- 2\P15\P25 + 2\P12\P34  + \P12\P35 + \P12\P45)ag_V\MCHSQ +
(X_{34} - Y_{34})bg_A +(- \P15\P23 - \P15\P24 + \P13\P25 +
 \P14\P25)bg_A\MCHSQ$
\vskip 2pt
$I_{38}=ag_A(\P34 + 2\MCHSQ)(\epsilon{(5124)} - \epsilon{(5123)}) + 
bg_V(\P23 - \P24)\epsilon{(5134)} + bg_V(\P13 - \P14)\epsilon{(5234)} + 
ag_A(-\P35 - \P45)\epsilon{(1234)}$
\end{flushleft} 
\begin{flushleft}
${\bf T_{45}}= 32Z_1Z_3\{S_1R_{45} - M_Z\GZ I_{45}\}/\P15/\P45$
\vskip 2pt
$R_{45}=\{ Y_{13}(3\leftrightarrow 4) + X_{13}(3\leftrightarrow 4)\}ag_V
 + (2\P12\P14 + \P12\P15 + \P15\P23 + \P15\P24 - 
 \P14\P25 - \P15\P25 - \P12\P45)ag_V\MCHSQ + 
\{ Y_{13}(3\leftrightarrow 4) - X_{13}(3\leftrightarrow 4)\}bg_A
+ (\P12\P15 + \P15\P23 - \P15\P25)bg_A\MCHSQ $
\vskip 2pt
$I_{45}=0.5\Big [\{ag_A(-\P14 + \MCHSQ) +
 bg_V (\P14 - \MCHSQ)\}\epsilon{(5123)} + 
\{ag_A( -P13 - P34 - 2\MCHSQ) + bg_V(\P13 + \P34)\}\epsilon{(5124)} +
 \{ag_A(-P12  - \P24) +bg_V(\P12 + \P24)\}\epsilon{(5134)} +
(-\P14ag_A + \P14bg_V)\epsilon{(5234)} +\{ag_A(-\P15 + P45) + bg_V(\P15 
 - \P45)\}\epsilon{(1234)} \Big ]$
\end{flushleft}
\begin{flushleft}
${\bf T_{46}}= -32Z_1Z_3\{S_1R_{46} - M_Z\GZ I_{46}\}/\P25/\P45$
\vskip 2pt
$R_{46}=\{X_{13}(1\leftrightarrow 2 \;\&\; 3
\leftrightarrow 4) +Y_{13}(1\leftrightarrow 2 \;\&\; 3\leftrightarrow 4)\}ag_V +
 (2\P12\P24 -  \P15\P24 + \P12\P25 + \P13\P25 + 
\P14\P25 - \P15\P25 - \P12\P45)ag_V\MCHSQ + \{X_{13}(1\leftrightarrow 2 \;\&\; 3
\leftrightarrow 4) - Y_{13}(1\leftrightarrow 2 \;\&\; 3\leftrightarrow 4)\}bg_A
 + (- \P12\P25 - \P13\P25 + \P15\P25)bg_A\MCHSQ $
\vskip 2pt
$I_{46}=0.5\Big [\{ag_A(-\P24 + \MCHSQ) 
bg_V(-\P24 + \MCHSQ)\}\epsilon{(5123)} + 
\{ag_A(-\P23 - \P34 - 2\MCHSQ) - bg_V(\P23 +\P34)\}\epsilon{(5124)} +
 (\P24ag_A + \P24bg_V)\epsilon{(5134)} + \{ag_A(\P12 + 
P14) + bg_V(\P12 + \P14)\}\epsilon{(5234)} + \{ag_A(-\P25 + 
P45) +b_GV(- \P25 + \P45)\}\epsilon{(1234)}\Big ]$
\end{flushleft}
\begin{flushleft}
${\bf T_{47}}= 16Z_1\{S_1R_{47} - M_Z\GZ I_{47}\}/\P12/\P35/\P45$
\vskip 2pt
$R_{47}= (X_{34} + Y_{34} - 2\P14\P24\P35 - 2\P13\P23\P45 )a_GV + 
(-2\P15\P25 + 2\P12\P34 + \P12\P35 + \P12\P45)ag_V\MCHSQ + (X_{34} -
 Y_{34})bg_A + \{\P15(\P23 + \P24) - (\P13 + \P14)\P25)\}bg_A\MCHSQ $
\vskip 2pt
$I_{47}=
ag_A(\P34 + 2\MCHSQ)\{\epsilon{(5123)} - \epsilon{(5124)}\} + 
bg_V\{(- \P23 + \P24)\epsilon{(5134)} + (- \P13 + \P14)\epsilon{(5234)}\} + 
ag_A(\P35 + \P45)\epsilon{(1234)}$
\end{flushleft}
\begin{flushleft}
${\bf T_{48}}= -32Z_1S_1R_{48}/\P12/\P45^2$
\vskip 2pt
$R_{48}= (ag_V + bg_A)\{-\P15\P23\P45 + \MCHSQ\P23(\P14 + \P15)\}
+ (ag_V - bg_A)\{-\P13\P25\P45 + \MCHSQ\P13(\P24 + \P25)\} +
 ag_V\MCHSQ\P12(\P45 + \MCHSQ)$
\end{flushleft}
\begin{flushleft}
GZ= $ov_2({\bf T_{15}} + {\bf T_{16}} + {\bf T_{17}} + {\bf T_{18}} +
{\bf T_{25}} + {\bf T_{26}}
 + {\bf T_{27}} + {\bf T_{28}} + {\bf T_{35}} + {\bf T_{36}} + {\bf T_{37}}
 + {\bf T_{38}} + {\bf T_{45}} +
 {\bf T_{46}} + {\bf T_{47}}+ {\bf T_{48}})$ 
\end{flushleft}
\vskip 5pt
\noindent
(4)Here we consider $s$-channel $\SNU$ exchange diagrams (9-12 in Fig. 1).
\begin{flushleft}
${\bf T_{99}}= 256\P23\P45/\P15/sn_2^2$
\end{flushleft}
\begin{flushleft}
${\bf T_{10(10)}}=256\P14\P35/\P25/sn_1^2$
\end{flushleft}
\begin{flushleft}
${\bf T_{11(11)}}=-256\P14\Big [-\P25\P35 +
 (P23 + P25)\MCHSQ\Big ]/\P35^2/sn_1^2$
\end{flushleft}
\begin{flushleft}
${\bf T_{12(12)}}=-256\P23\Big [-\P15\P45 +
 (P14 + P15)\MCHSQ\Big ]/\P45^2/sn2^2$
\end{flushleft}
\begin{flushleft}
${\bf T_{9(10)}}=256(X_{25} + \P15\P23\P24 + \P13\P14\P25)/\P15/\P25/sn_1/sn_2$
\end{flushleft}
\begin{flushleft}
${\bf T_{9(11)}}=-256X_{13}/\P15/\P35/sn_1/sn_2$
\end{flushleft}
\begin{flushleft}
${\bf T_{9(12)}}=256\Big [Y_{13}(3\leftrightarrow 4) +
 \MCHSQ\P15\P23\Big ]/\P15/\P45/sn_2^2$
\end{flushleft}
\begin{flushleft}
${\bf T_{10(11)}}=256\Big [Y_{13}(1\leftrightarrow 2) +
 \MCHSQ\P14\P25\Big ]/\P25/\P35/sn_1^2$
\end{flushleft}
\begin{flushleft}
${\bf T_{10(12)}}=-256\Big [X_{13}(1\leftrightarrow 2 \;\&\; 3\leftrightarrow 4)
\Big ]/\P25/\P45/sn_1/sn_2$
\end{flushleft}
\begin{flushleft}
${\bf T_{11(12)}}=256\{X_{34} - \P14\P24\P35 -
 \P13\P23\P45\} /\P35/\P45/sn_1/sn_2$
\end{flushleft}
\begin{flushleft}
SS= $ov_4({\bf T_{99}} + {\bf T_{10(10)}} + {\bf T_{11(11)}} + {\bf T_{12(12)}}
 + {\bf T_{9(10)}} + {\bf T_{9(11)}} + {\bf T_{9(12)}}
 + {\bf T_{10(11)}} + {\bf T_{10(12)}} + {\bf T_{11(12)}})$        
\end{flushleft}
\noindent
(SS:represents the total $\SNU$-exchange matrix element squard)
\vskip 5pt
\noindent
(5)Interference between $s$ (Z-exchange) and $t$ ($\SNU$-exchange) channel 
diagrams is considered here. 
\begin{flushleft}
${\bf T_{19}}=128Z_2S_2\{(a+b)(g_V - g_A)(2\P23\P45 + \MCHSQ\P25 \}/sn_2/\P15$
\end{flushleft}
\begin{flushleft}
${\bf T_{1(10)}}=-128Z_2\{S_2R_{1(10)} - \GZ M_ZI_{1(10)}\}/sn_1/\P15/\P25$
$R_{1(10)}=(g_A - g_V)\Big [ (a - b)(X_{25} + \P15\P23\P24 +
 \P13\P14\P25) + \MCHSQ(a + b)\P12(\P12 -\P15 - \P25)\Big ] $
$I_{1(10)}=0.5(g_V - g_A)(a - b)\Big [2(\P14 - \P45)\epsilon{(5123)} +
(\P35 - \P23)\epsilon{(5124)} - \P25\epsilon{(5134)} +
\P15\epsilon{(5234)}\Big ]$
\end{flushleft}
\begin{flushleft}
${\bf T_{1(11)}}=64Z_2\{S_2R_{1(11)} - \GZ M_ZI_{1(11)}\}/sn_1/\P15/\P35$
$R_{1(11)}=(g_A - g_V)\Big [ 2(a-b)X_{13} + \MCHSQ(2\P12\P13 + \P15\P23
 - \P13\P25 - \P12\P35) \Big ]$
$I_{1(11)}=(g_V - g_A)\Big [(b-a)\{ -\P14\epsilon{(5123)} + (\MCHSQ -
\P13)\epsilon{(5124)} + (\P12 + \P23)\epsilon{(5134)} + \P13\epsilon{(5234)}
+ (\P15 - \P35)\epsilon{(1234)}\} + (a +b)\epsilon{(5123)}\Big ]$
\end{flushleft}
\begin{flushleft}
${\bf T_{1(12)}}=-64Z_2\{S_2R_{1(12)} - \GZ M_ZI_{1(12)}\}/sn_2/\P15/\P45$
$R_{1(12)}=(g_A - g_V)\Big [ 2(a-b)Y_{13}(3\leftrightarrow 4) +
\MCHSQ N_{1(12)}\Big ]$
$N_{1(12)}=2\P12\P14 + 2\P12\P15 + 2\P15\P23 + \P15\P24 - 
 \P14\P25 - 2\P15\P25 - \P12\P45$
$I_{1(12)}=\MCHSQ(a+b)(g_V - g_A)\epsilon{(5124)}$
\end{flushleft}
\begin{flushleft}
${\bf T_{29}}=-128Z_2\{S_2R_{29} - \GZ M_ZI_{29}\}/sn_2/\P15/\P25$
$R_{29}=(g_A - g_V)\Big [ (a-b)\{X_{25} + \P15\P23\P24 + \P13\P14\P25\}
+\MCHSQ(a+b)\{\P12(\P12 -\P15 -\P25)\}\Big ]$
$I_{29}=0.5(g_V - g_A)(b-a)\Big [ (\P14 - \P24)\epsilon{(5123)} +(\P13 -
\P23)\epsilon{(5124)} + \P12\{\epsilon{(5234)} - \epsilon{(5134)}\} 
-(\P15 + \P25)\epsilon{(1234)}\Big ]$
\end{flushleft}
\begin{flushleft}
${\bf T_{2(10)}}=128Z_2S_2(g_V - g_A)\Big [ 2(a-b)\P14\P25\P35 
+ \MCHSQ(a+b)\P15\P25\Big ]/sn1/\P25^2$
\end{flushleft}
\begin{flushleft}
${\bf T_{2(11)}}=-64Z_2\{S_2R_{2(11)} - \GZ M_ZI_{2(11)}\}/sn_1/\P25/\P35$
$R_{2(11)}=(g_A - g_V)\Big [ 2(a-b)Y_{13}(1\leftrightarrow 2) + 
\MCHSQ(a+b)N_{1(12)}(1\leftrightarrow 2 \;\&\; 3\leftrightarrow 4)\Big ]$
$I_{2(11)}=\MCHSQ(a+b)(g_V - g_A)\epsilon{(5123)}$
\end{flushleft}
\begin{flushleft}
${\bf T_{2(12)}}=64Z_2\{S_2R_{2(12)} - \GZ M_ZI_{2(12)}\}/sn_2/\P25/\P45$\\
$R_{2(12)}=(g_A - g_V)\Big [ 2(a-b)X_{13}(1\leftrightarrow 2 \;\&\;
3\leftrightarrow 4) +\MCHSQ(a+b)(2\P12\P24 - \P15\P24 + \P14\P25 - \P12\P45)
\Big ]$\\
$I_{2(12)}=(g_V - g_A)\Big [ (a-b)\{ \P24\epsilon{(5123)} + 
\P23\epsilon{(5124)} - (\P12 + 2\P14)\epsilon{(5234)} + \P25\epsilon{(1234)}\}
+ \MCHSQ(a+b)\epsilon{(5124)}\Big ]$
\end{flushleft}
\begin{flushleft}
${\bf T_{39}}=-64Z_1\{S_1R_{39} - \GZ M_ZI_{39}\}/sn_2/\P15/\P35$\\
$R_{39}=(g_A - g_V)\Big [ 2(b-a)X_{13} - \MCHSQ(a+b)(2\P12\P13 +
  2\P12\P15 + \P15\P23 - \P13\P25 - 2\P15\P25 - \P12\P35)\Big ]$
$I_{39}=(g_V - g_A)\Big [ (a-b)\{(\P14 + \P34)\epsilon{(5123)} +
\P13(\epsilon{(5124)} - \epsilon{(5234)})- (\P12 + \P23)\epsilon{(5134)}-
(\P35 - \P15)\epsilon{(1234)}\} + \MCHSQ\{ (a+b)\epsilon{(5123)} +(b-a)
\epsilon{(5124)}\} \Big ]$
\end{flushleft}
\begin{flushleft}
${\bf T_{3(10)}}=64Z_1\{S_1R_{3(10)} - \GZ M_ZI_{3(10)}\}/sn_1/\P25/\P35$\\
$R_{3(10)}=(g_A - g_V)\Big [4(b-a)\P14(\P23^2 + \P23\P25 - \P23\P35 -
 \P25\P35) - (a+b)\MCHSQ(2\P12\P13 +  2\P12\P15 + \P15\P23 - \P13\P25
 -  2\P15\P25 - \P12\P35)\Big ]$\\
$I_{3(10)}=(g_V - g_A)(a+b)\epsilon{(5123)}$
\end{flushleft}
\begin{flushleft}
${\bf T_{3(11)}}= -128Z_1S_1(g_A-g_V)\Big [2(a-b)\P14\{\P25\P35
 -\MCHSQ(\P23 + \P25)\} -\MCHSQ(a+b)\P12\{\MCHSQ +\P35 \} \Big ]/sn_1/\P35^2$
\end{flushleft}
\begin{flushleft}
${\bf T_{3(12)}}= 64Z_1\{S_1R_{3(12)} - \GZ M_Z I_{3(12)}\}/sn_2/\P35/\P45$
$R_{3(12)}=(g_A - g_V)\Big [2(b-a)\{X_{34} - \P14\P24\P35-\P13\P23\P45\}
-\MCHSQ(a+b) \{\P15\P23 + \P15\P24 - \P13\P25 - \P14\P25 - 2\P15\P25 + 
 2\P12\P34 + \P12\P35 + \P12\P45)\}\Big ]$
$I_{3(12)}=(g_V - g_A)\Big [(a-b)\{\P34(\epsilon{(5124)}-
\epsilon{(5123)}) + (\P24 - \P23)\epsilon{(5134)} + (\P14 -
 \P13)\epsilon{(5234)} + (\P35 +\P45)\epsilon{(1234)}\} + 2a\MCHSQ\{
\epsilon{(5123)} - \epsilon{(5124)}\}\Big ]$
\end{flushleft}
\begin{flushleft}
${\bf T_{49}}= 64Z_1\{S_1R_{49} - \GZ M_ZI_{49}\}/sn_2/\P15/\P45$

$R_{49}=(g_A - g_V)\Big [2(b-a)\{Y_{13}(3\leftrightarrow 4)\} -
\MCHSQ(a+b)(2\P12\P14 + 2\P15\P23 + \P15\P24 - \P14\P25 - \P12\P45)\Big ]$
$I_{49}=(a+b)(g_V - g_A)\MCHSQ\epsilon{(5124)}$
\end{flushleft}
\begin{flushleft}
${\bf T_{4(10)}}= -64Z_1\{S_1R_{4(10)} - \GZ M_ZI_{4(10)}\}/sn_1/\P25/\P45$\\
$R_{4(10)}=(g_A - g_V)\Big [2(b-a)\{X_{13}(1\leftrightarrow 2 \;\&\;
3\leftrightarrow 4)\} - \MCHSQ(a+b)\{2\P12\P24 - \P15\P24 +
 2\P12\P25 + \P14\P25 - 2\P15\P25 - \P12\P45\}\Big ]$\\
$I_{4(10)}=(g_V - g_A)\Big [ (a-b)\{(2\P24 - \MCHSQ)\epsilon{(5123)}
+ \P34\epsilon{(5124)} - \P24\epsilon{(5134)} - (2\P12 + \P14)\epsilon{(5234)}
- (\P45 - 2\P25)\epsilon{(1234)}\} + \MCHSQ(a+b)\epsilon{(5124)} \Big ]$
\end{flushleft}
\begin{flushleft}

${\bf T_{4(11)}}= 64Z_1\{S_1R_{4(11)} - \GZ M_ZI_{4(11)}\}/sn_1/\P35/\P45$\\

$R_{4(11)}=(g_A - g_V)\Big [2(b-a)\{X_{34} - \P14\P24\P35 - 
\P13\P23\P45\} + \MCHSQ(a+b)\{\P15\P23 + \P15\P24 - \P13\P25 -
\P14\P25 + 2\P15\P25 - 2\P12\P34 - \P12\P35 - \P12\P45\}\Big ]$

$I_{4(11)}=(g_V - g_A)\Big [(a-b)\{\P34(\epsilon{(5124)} -
 \epsilon{(5123)} + (\P23 - \P24)\epsilon{(5134)} + (\P13 -
 \P14)\epsilon{(5234)} - (\P35 + \P45)\epsilon{(1234)} \} +2a\MCHSQ\{
\epsilon{(5124)} - \epsilon{(5123)}\} \Big ]$
\end{flushleft}
\begin{flushleft}
${\bf T_{4(12)}}= -128Z_1S_1(g_A - g_V)\Big [2(a-b)\P23\{\P15\P45
 -\MCHSQ(\P14 + \P15)\} -\MCHSQ(a+b)\P12\{\MCHSQ +\P45 \} \Big ]/sn_2/\P45^2$
\end{flushleft}
\begin{flushleft}
ZS= $ov_5({\bf T_{19}} + {\bf T_{1(10)}} + {\bf T_{1(11)}} +{\bf T_{1(12)}} +
{\bf T_{29}} + {\bf T_{2(10)}} + {\bf T_{2(11)}} + {\bf T_{2(12)}}  +
 {\bf T_{39}} + {\bf T_{3(10)}} + {\bf T_{3(11)}} + {\bf T_{3(12)}} + 
{\bf T_{49}} + {\bf T_{4(10)}} + {\bf T_{4(11)}} + {\bf T_{4(12)}})$       
\end{flushleft}
\vskip 5pt
\noindent
(6)Interference between $s$ ($\gamma$-exchange) and $t$ ($\SNU$-exchange) channel 
diagrams is considered here. 
\begin{flushleft}
${\bf T_{59}}= 128Z_3\{2\P23\P45 + \P25\MCHSQ\}/sn_2/\P15$
\end{flushleft}
\begin{flushleft}
${\bf T_{5(10)}}= 128Z_3\Big [ X_{25}  + \P15\P23\P24 + \P13\P14\P25 +
 \MCHSQ\P12(\P12 - \P15 - \P25)\Big ]/sn_1/\P15/\P25$
\end{flushleft}
\begin{flushleft}
${\bf T_{5(11)}}=-64Z_3\Big [ 2X_{13} + (2\P12\P13 +  \P15\P23
 - \P13\P25 - \P12\P35)\MCHSQ\Big ]/sn_1/\P15/\P35$
\end{flushleft}
\begin{flushleft}
${\bf T_{5(12)}}=64Z_3\Big [ 2Y_{13}(3\leftrightarrow 4) +
 \MCHSQ N_{1(12)}\Big ]/sn_2/\P15/\P45$
\end{flushleft}
\begin{flushleft}
${\bf T_{69}}= 128Z_3\Big [ X_{25}  + \P15\P23\P24 + \P13\P14\P25 +
 \MCHSQ\P12(\P12 - \P15 - \P25)\Big ]/sn_2/\P15/\P25$
\end{flushleft}
\begin{flushleft}
${\bf T_{6(10)}}=128Z_3\{2\P14\P35 + \P15\MCHSQ\}/SN1/\P25$
\end{flushleft}
\begin{flushleft}
${\bf T_{6(11)}}=64Z_3\Big [ 2Y_{13}(1\leftrightarrow 2) +
 \MCHSQ N_{1(12)}(1\leftrightarrow 2 \;\&\; 3\leftrightarrow 4)
 \Big ]/sn_1/\P25/\P35$
\end{flushleft}
\begin{flushleft}
${\bf T_{6(12)}}=-64Z_3\Big [2X_{13}(1\leftrightarrow 2 \;\&\; 3\leftrightarrow 4)
 + (2\P12\P24 - \P15\P24 + \P14\P25 - \P12\P45)\MCHSQ \Big ] /sn_2/\P25/\P45$ 
\end{flushleft}
\begin{flushleft}
${\bf T_{79}}=-32\Big [ 2X_{13} + \MCHSQ\{N_{1(12)}(3\leftrightarrow 4)-
2\P15\P24\}\Big ]/sn_2/\P12/\P15/\P35$
\end{flushleft}
\begin{flushleft}
${\bf T_{7(10)}}=32\Big [2Y_{13}(1\leftrightarrow 2) + (2\P12\P23 -
 \P15\P23 + \P13\P25 + 2\P14\P25 - \P12\P35)\MCHSQ\Big ]/sn_1/\P12/\P25/\P35$
\end{flushleft}
\begin{flushleft}
${\bf T_{7(11)}}= -64\Big [-2\P14\P25\P35 + \{2\P14\P23 +
 2\P14\P25 + \P12\P35 + \P12\MCHSQ\}\MCHSQ \Big ]/sn_1/\P12/\P35^2$
\end{flushleft}
\begin{flushleft}
${\bf T_{7(12)}}=32\Big [2(X_{34} - \P14\P24\P35 - \P13\P23\P45) +
\MCHSQ (\P15\P23 +  \P15\P24 - \P13\P25 - \P14\P25 - 2\P15\P25 + 
 2\P12\P34 + \P12\P35 + \P12\P45) \Big ]/sn_2/\P12/\P35/\P45$
\end{flushleft}
\begin{flushleft}
${\bf T_{89}}=32\Big [2Y_{13}(3\leftrightarrow 4) + \MCHSQ(2\P12\P14 +
 2\P15\P23 + \P15\P24 - \P14\P25 - \P12\P45)\Big ]/sn_2/\P12/\P15/\P45$
\end{flushleft}
\begin{flushleft}
${\bf T_{8(10)}}=-32\Big [2X_{13}(1\leftrightarrow 2\;\&\; 3\leftrightarrow 4) +
\MCHSQ(2\P12\P24 - \P15\P24 + 2\P12\P25 + \P14\P25 -
2\P15\P25 - \P12\P45) \Big ]/sn_1/\P12/\P25/\P45$
\end{flushleft}
\begin{flushleft}
${\bf T_{8(11)}}=32\Big [2(X_{34} -\P14\P24\P35 - \P13\P23\P45) +
\MCHSQ(-\P15\P23 - \P15\P24 + \P13\P25 + \P14\P25 -2\P15\P25 + 2\P12\P34
 + \P12\P35 +\P12\P45) \Big ]/sn_1/\P12/\P35/\P45$
\end{flushleft}
\begin{flushleft}
${\bf T_{8(12)}}= -64\Big [- 2\P15\P23\P45 + \{2\P14\P23 +
2\P15\P23 + \P12\P45 + \P12\MCHSQ\}\MCHSQ \Big ] /sn_2/\P12/\P45^2$
\end{flushleft}
\begin{flushleft}
SG= $ov_6( {\bf T_{59}} + {\bf T_{5(10)}} + {\bf T_{5(11)}} + {\bf T_{5(12)}}
 + {\bf T_{69}} + {\bf T_{6(10)}} + {\bf T_{6(11)}} + {\bf T_{6(12)}}  +
 {\bf T_{79}} + {\bf T_{7(10)}} + {\bf T_{7(11)}} + {\bf T_{7(12)}} + 
{\bf T_{89}} + {\bf T_{8(10)}} + {\bf T_{8(11)}} + {\bf T_{8(12)}})$           
\end{flushleft}
\begin{flushleft}
Total matrix element squared:\\
MSQ= ZZ + GG + GZ + SS + SZ + SG
\end{flushleft}
\begin{flushleft}
The differential cross section is given by
\[ d \sigma = S_{avg} \frac{(MSQ)}
{64 E_{CM}^2 \pi ^5}  \delta ^4 \Big ( p_1+p_2- \sum_{i=3}^5
p_i \Big ) \prod_{i=3}^5 \frac{d^3p_i}{2E_i} \]
\end{flushleft}
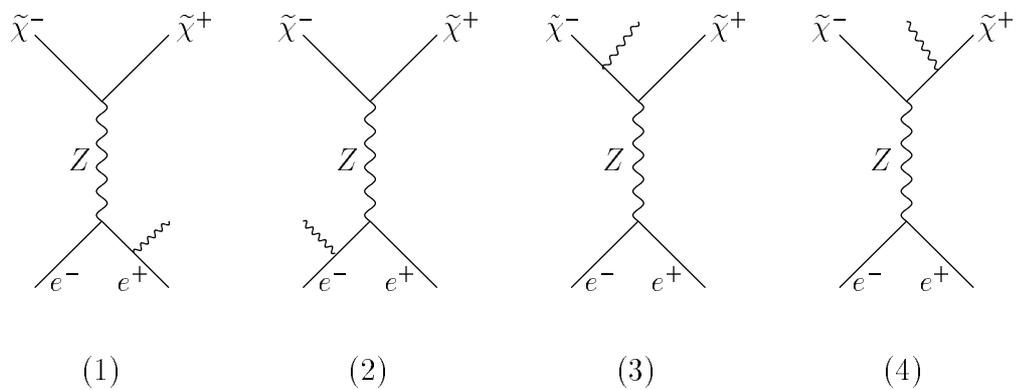
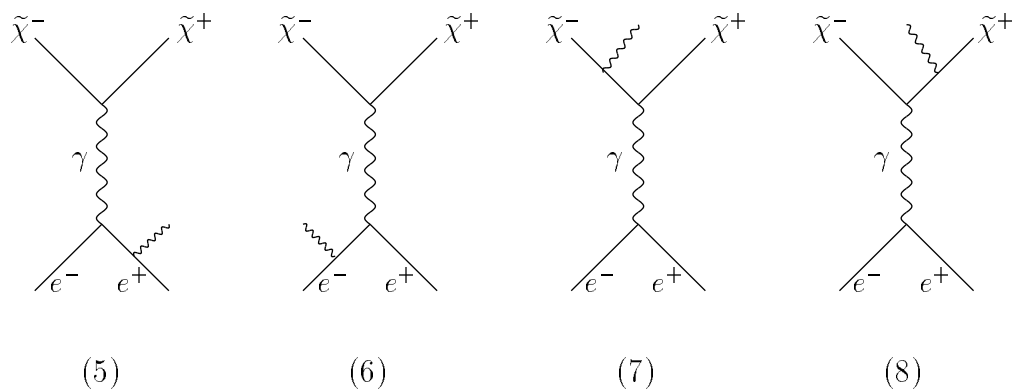
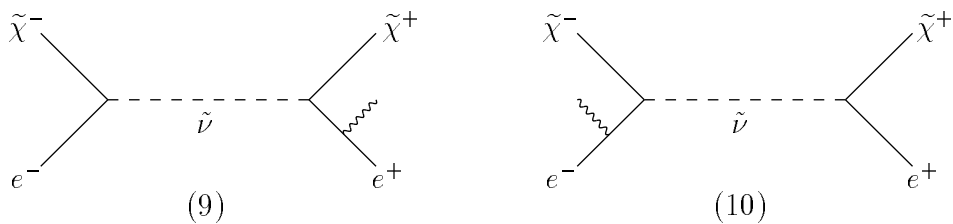
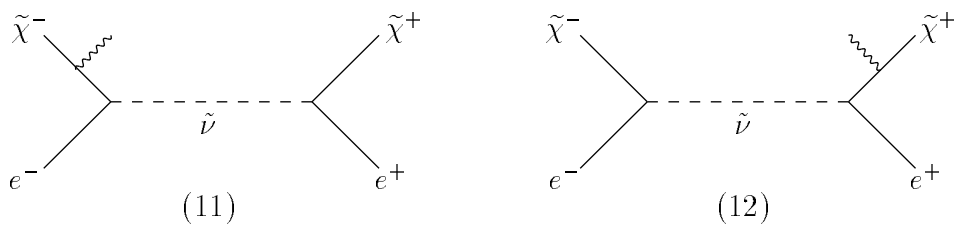
\end{document}